\documentstyle[11pt]{article}
\newcommand{\reell}{{\kern+.25em\sf{R}\kern-.78em\sf{I}\kern+.78em\kern-.25em}}
\newcommand{\natur}{{\kern+.25em\sf{N}\kern-.86em\sf{I}\kern+.86em\kern-.25em}}

\begin{document}
\begin{center}
{\Large\bf Solutions of the Spherically Symmertic \\
\vspace*{2mm}
Wave Equation in $p+q$ Dimensions}
\footnote{Work supported by Conselho Nacional de Desenvolimento
Cientifico e Tecnologico (CNPq)}

\vspace*{1cm}

W. Bietenholz \ and  \ J.J. Giambiagi \\
Centro Brasileiro de Pesquisas Fisicas (CBPF)\\
Rua Dr. Xavier Sigaud 150 \\
22290-180 Rio de Janeiro, RJ \\
Brazil\\

\end{center}

{\em Abstract} \ We discuss solutions of the
spherically symmetric wave equation and Klein Gordon equation
in an arbitrary number of spatial and temporal dimensions.
Starting from a given solution,
we present various procedures to generate further
solutions in the same or in different dimensions.
The transition from odd to even and non
integer dimensions can be performed by fractional derivation
or integration. The dimensional shift, however, can also be
interpreted simply as a modification of the dynamics.
We also discuss the analytic continuation to arbitrary real powers of
the D'Alembert operator. There, particular peculiarities in the pole
structure show up when $p$ and $q$ are both even.
Finally, we give operators which transform a time into a
space coordinate and v.v. and comment on their possible relation
to black holes. In this context, we describe a few aspects of the
extension of our discussion to a curved metrics.

\section{Introduction}

The wave equation plays a very important role in practically all branches of
physics. It has a fundamental meaning in classical as well as
quantum physics, including field theory. This refers to both,
the non relativistic as well as the relativistic description. Hence it
is strongly motivated to discuss solutions of the wave eq.,
or variants thereof such as the Klein Gordon eq., in all possible
situations.

Also the use of space-time dimensions different from $3+1$
is well established nowadays in most parts of physics. In this
paper we provide a
background of solutions for the spherically symmetric wave eq.,
which may turn out to be useful anywhere. But we do not discuss
particular applications here.

Recently, Bollini and Giambiagi described fractional powers of the
D'Alembert operator and of the Laplace operator in one temporal
and an arbitrary number of spatial dimensions and discussed Huygens'
principle and causality in this framework \cite{BG}. In a subsequent
paper, Giambiagi referred to the standard D'Alembert operator
and discussed relations among solutions of the wave
eq. again in one temporal and different spatial dimensions. He also
considered the Klein Gordon eq. and Green's functions.

Now we want to generalize both of these considerations
to $p+q$ dimensions. We are particularly interested in operations
shifting the dimensions in which a solution is valid.
As we will see, interesting and qualitatively new
properties arise, particularly in the pole structure for even $p$ and $q$.
The mathematical background for this is beautifully outlined in the
classical book of Gelfand and Shilov \cite{GS}.
At the end, we add some remarks about a generalization to curved metrics.
This is motivated from the fact that transitions of a spatial
to a temporal dimension and vice versa are known in general relativity:
if we cross the boundary of a black hole -- described e.g. by the
Schwarzschild metrics -- time becomes spatial and
the radial component becomes temporal.

\section{The spherical wave equation in $d+1$ dimension}

The case of one time and an arbitrary number of spatial
dimensions has been discussed extensively in \cite{jjg}.
We start by adding some complementary observations to this
case.

We consider solutions of the spherically symmetric wave eq.
\begin{equation} \label{wed1}
\Big[ \partial_{r}^{2} + \frac{d-1}{r} \partial_{r} - \partial_{t}^{2} \Big]
\phi_{d}(t,r) = 0
\end{equation}
where $\partial_{r} \doteq \partial / \partial r $ etc. We are not
interested in constant factors or additive constants in $\phi_{d}$.

With the well known substitution:
\begin{equation}
\Omega_{d}(t,r) \doteq r^{(1-d)/2} \phi_{d}(t,r)
\end{equation}
we can absorb the linear derivative. Eq. (\ref{wed1}) takes the form:
\begin{equation}
\Big[ \partial_{r}^{2} + \frac{(d-1)(d-3)}{4r^{2}} - \partial_{t}^{2} \Big]
\Omega_{d}(t,r) =0
\end{equation}
and we can immediately read off general solutions for $d \in \{1,3 \}$:
\begin{eqnarray}
\Omega_{1}(t,r) &= F_{1}(t-r) + F_{2}(t+r) \ ; \quad
\Omega_{3}(t,r) &= f_{1}(t-r) + f_{2}(t+r) \nonumber \\
\to \quad \phi_{1}(t,r) &= F_{1}(t-r)+F_{2}(t+r) \ ; \quad
\phi_{3}(t,r) &= \frac{f_{1}(t-r)+f_{2}(t+r)}{r} \label{d13}
\end{eqnarray}
where $F_{1},F_{2},f_{1},f_{2}$ are arbitrary functions or symbolic functions.
If we choose $\delta $ functions for them, the above solutions represent
waves which are: for $d=1$ moving to the right/left, and for $d=3$
outgoing/incoming.

Here $d=1$ and $d=3$ play a special role, since for no other dimension
there is a solution of the type:
\begin{equation}
\phi_{d}(t,r) = f(t \ ^{+}_{-} \ r)r^{\alpha} \qquad (\alpha \in \reell )
\end{equation}
This can be seen by inserting this ansatz in (\ref{wed1}), which
yields the conditions: $2 \alpha = 1-d $ and $ \alpha^{2}=\alpha
(2-d)$, with the only solutions: $ \alpha =0, \ d=1 $ or $\alpha
=-1, \ d=3$.

In \cite{jjg} it was shown that if $\phi_{d}$ is a solution of (\ref{wed1}),
then the same holds for
\begin{equation} \label{difd}
\phi_{d+2}(t,r) = - \frac{1}{r} \partial_{r} \phi_{d}(t,r)
\end{equation}
With this respect, we can identify in (\ref{d13}): $F_{1}'=f_{1}, \
F_{2}'=f_{2}$ and we obtain a set of solutions for all odd $d$,
generated by the choice for $F_{1}, \ F_{2}$, e.g.
\begin{equation}
\phi_{5}(t,r) = \frac{f_{1}(t-r)+f_{2}(t+r)}{r^{3}} +
\frac{f_{1}'(t-r)-f_{2}'(t+r)}{r^{2}} \qquad {\rm etc.}
\end{equation}
Let's consider the two signs individually and first let $F_{2} \equiv 0$.
Then the solutions for all odd $d$ -- generated from $F_{1}$ by means
of (\ref{difd}) -- have the form:
\begin{equation} \label{series}
\phi_{2k+1} (t,r) = \sum_{n=0}^{k} a_{n}^{(k)} \frac{f^{(n)}(t-r)}
{r^{2k-n-1}}
\end{equation}
The coefficients $a_{n}^{(k)}$ are the elements of a ``modified
Pascalian triangle'':
\begin{equation} \label{pascal}
\begin{array}{ccccccccccccc}
k=0 &&&&&& 1 &&&&&& \\
k=1 &&&&&& 1 & 1 &&&&& \\
k=2 &&&&&& 3 & 3 & 1 &&&& \\
k=3 &&&&&& 15 & 15 & 6 & 1 &&& \\
k=4 &&&&&& 105 & 105 & 45 & 10 & 1 &&\\
k=5 &&&&&& 945 & 945 & 420 & 105 & 15 & 1 &\\
\dots &&&&&& \dots &&&&& \\
&&&&&& n=0 & n=1 & n=2 & n=3 & n=4 & n=5  & \dots
\end{array}
\end{equation}
($d=1$ is not represented here.)

The elements at the margins are: $a_{0}^{(k)}=(2k-1)!! \ , \ a_{k}^{(k)} =1$,
and the rest is determined by the recursion formula:
\begin{equation} \label{recur}
a_{n}^{(k)} = a_{n-1}^{(k-1)}+(2k-n-1)a_{n}^{(k-1)}
\end{equation}
As in the Pascalian triangle, the elements off the margin are built
from the two elements vertically and on the left above. The only
modifications are the left margin and the factor $(2k-n-1)$ in
(\ref{recur}).

We add two observations for the elements next to the margins, which
can easily be proved by induction. For all $k \in \natur $ these elements
take the form:
\begin{equation}
a_{1}^{(k)} = a_{0}^{(k)} \ , \qquad a_{k-1}^{(k)}
= \left( \begin{array}{c} k+1 \\ 2 \end{array} \right)
\end{equation}

Now we consider the other case: $F_{1} \equiv 0$. The solutions
generated by (\ref{difd}) from $F_{2}$ can be represented in
the same way as (\ref{series}):
\begin{equation}
\phi_{2k+1}(t,r) = \sum_{n=0}^{k} \bar a_{n}^{(k)} \frac{f^{(n)}(t+r)}
{r^{2k-n-1}}
\end{equation}
The triangle for $\bar a_{n}^{(k)}$, however, is different:
\begin{equation}
\begin{array}{ccccccccccccc}
k=0 &&&&&& 1 &&&&&& \\
k=1 &&&&&& 1 & -1 &&&&& \\
k=2 &&&&&& 3 & -1 & 1 &&&& \\
k=3 &&&&&& 15 & -1 & 2 & -1 &&& \\
k=4 &&&&&& 105 & 9 & 9 & -2 & 1 && \\
k=5 &&&&&& 945 & 177 & 72 & -3 & 3 & -1 & \\
\dots &&&&&& \dots &&&&&& \\
&&&&&& n=0 & n=1 & n=2 & n=3 & n=4 & n=5 & \dots \\
\end{array}
\end{equation}
The left margin is the same as in (\ref{pascal}) and also the
recursion relation (\ref{recur}) still holds, but the right margin
is oscillating: $\bar a_{k}^{(k)} = (-1)^{k}$, thus altering all the
off margin elements. For the second column from the right we have the
general expression:
\begin{equation}
\bar a_{k-1}^{(k)} = \Big\{ \begin{array}{ccc} (k+1)/2 && k {\rm ~~odd} \\
- k/2 && k {\rm ~~even} \end{array}
\end{equation}

\section{The spherical wave equation in $p+q$ dimensions}

We consider a flat space with coordinates $(t_{1}, \dots , t_{q},
x_{1}, \dots x_{p})$ and search for solutions of the wave eq., which
depend only on $r \doteq \sqrt{\sum_{i=1}^{p} x_{i}^{2}}$ and
$\tau \doteq \sqrt{\sum_{i=1}^{q}t_{i}^{2}}$. They have to fulfill:
\begin{equation} \label{we}
\Big[ \partial_{r}^{2}+ \frac{p-1}{r}\partial_{r}-\partial_{\tau}^{2}
- \frac{q-1}{\tau} \partial_{\tau} \Big] \phi_{p,q}(\tau ,r) = 0
\end{equation}
i.e. we generalize the case $q=1$ considered in section 1.
Accordingly, we also generalize the first ansatz to:
\begin{equation} \label{ansa}
\phi_{p,q}(\tau ,r) = f(\tau \ ^{+}_{-} \ r) \tau^{\alpha}r^{\beta}
\qquad (\alpha ,\beta \in R)
\end{equation}
Again the wave eq. imposes constraints, which allow $\alpha $ and $\beta $
only to take the values $0$ or $-1$, and therefore: $p,q \in \{ 1,3 \}$.
The only new solution that we obtain is:
\begin{equation} \label{33}
\phi_{3,3}(\tau ,r) = \frac{f(\tau \ ^{+}_{-} \ r)}{\tau r}
\end{equation}
Now we search again for a rule how to generate solutions in higher
dimensions starting from these particular ones.\\

{\em\bf Rule 1} If $\phi_{p,q}$ is a solution of the wave eq.
(\ref{we}), then also
\begin{displaymath}
\phi_{p+2,q} \doteq - \frac{1}{r} \partial_{r} \phi_{p,q} \quad
{\rm and} \quad \phi_{p,q+2} \doteq - \frac{1}{\tau} \partial_{\tau}
\phi_{p,q}
\end{displaymath}
are solutions (always for the corresponding dimensions). \\

This generalizes the rule (\ref{difd}) and yields for example
the solution (\ref{33}). The proof by induction is straightforward,
if we just consider
\begin{displaymath}
\Big[ \partial^{2}_{r}+\frac{p+1}{r} \partial_{r} \Big] \frac{1}{r}
\partial_{r} = \frac{1}{r} \partial_{r} \Big[ \partial^{2}_{r}
+ \frac{p-1}{r} \partial_{r} \Big]
\end{displaymath}
and the same for $\tau $.

Hence starting from any $\phi_{1,1} = F_{1}(\tau -r)+F_{2}(\tau +r)$
we can generate immediately solutions {\em for all odd} $p$ and $q$.
If we choose $F_{1},F_{2}$ to be step functions, then the degree
of the pole in a solution $\phi_{p,q}$, which is generated
by application of rule 1, is $(p-1)/2 +(q-1)/2$. In particular
this set of solutions includes the physical monopole solution in
3+1 dimensions.

Now we want to postulate a second rule for generating solutions
in higher dimensions from a given one:

{\em\bf Rule 2} If $\phi_{p,q}$ is a solution of (\ref{we}),
then also
\begin{displaymath}
\bar \phi_{p+2,q+2} \doteq - \Big[ \frac{1}{r} \partial_{r} + \frac{1}{\tau}
\partial_{\tau} \Big] \phi_{p,q}
\end{displaymath}
is a solution.\\

The proof consists just of inserting rule 1.
But we emphasize that $\bar \phi_{p+2,q+2}$ does {\em not} coincide with
$\phi_{p+2,q+2} = \frac{1}{r \tau } \partial_{r} \partial_{\tau} \phi_{p,q}$.
To clarify the situation, we introduce the concept of equivalence
classes of solutions in different dimensions:

{\em Definition:} The solutions $\phi_{p+2,q} \ (\phi_{p,q+2})$ and
$\phi_{p,q}$ belong to the same equivalence class, if they are related
as $\phi_{p+2,q} = -\frac{1}{r} \partial_{r} \phi_{p,q}$
($\phi_{p,q+2} = - \frac{1}{\tau} \partial_{\tau} \phi_{p,q}$),
where we really mean ``equal to'', excluding different additive or
multiplicative constants.\\

Hence every solution $\phi_{1,1}$ defines an equivalence class with a unique
$\phi_{p,q}$ for all odd $p,q$. Now rule 2 can be formulated like this:

If $\phi_{p+2,q}$ and $\phi_{p,q+2}$ are solutions belonging to the
same equivalence class, then their superposition
$\bar \phi_{p+2,q+2} = \phi_{p+2,q}+\phi_{p,q+2}$ is a solution too,
which does, however, not belong to the same equivalence class.

Of course we may also define a second type of equivalence classes
of solutions related by rule 2. In such classes, $p-q$ must be fix.
Starting from a given $\phi_{p,q}$, all
\begin{equation}
\bar \phi_{p+2n,q+2n} = \sum_{k=0}^{n} (^{n}_{k}) \phi_{p+2k,q+2(n-k)}
\end{equation}
belong to the same second type equivalence class as
$\phi_{p,q}$, if all the $\phi_{p+2k,q+2(n-k)}$ belong to the same
first type equivalence class as $\phi_{p,q}$. This generalizes the above
formulation.\\

Let's reconsider $p=q=3$ and start from $\phi_{1,1}$ given in
(\ref{d13}). We obtain:
\begin{eqnarray}
\phi_{3,3}(\tau ,r) &=& \frac{1}{\tau r} [-F_{1}''(\tau -r)+F_{2}''(\tau + r)]
\nonumber \\
\bar \phi_{3,3}(\tau ,r) &=& \frac{1}{\tau r} [ (\tau -r)F_{1}'
(\tau -r) + (\tau + r) F_{2}'(\tau + r)]
\end{eqnarray}
These functions are different, but not basically different in the
sense that both of them fit with solution (\ref{33}).

If we choose $F_{1},F_{2}$ to be step functions, then $F_{1}'',F_{2}''=
\delta '$, i.e. we get from the first rule a dipole solution in
3+3 dimensions. (We noted before that in general the pole has the degree
$(p/2 + q/2 -1)$). The second rule seems to generate a monopole solution.
But in fact it vanishes since $x \delta (x) \equiv 0$.\\

To see that the two prescriptions do provide basically different solutions
in general, we consider as an example
$p=q=5$ and start from $\phi_{3,3} = f(\tau -r)/\tau r$:
\begin{eqnarray}
\phi_{5,5}(\tau ,r) &=& \frac{1}{(\tau r)^{3}} [ f(\tau -r) - f'(\tau -r)
(\tau -r) - f''(\tau -r)\tau r] \\
\bar \phi_{5,5}(\tau r) &=& \frac{1}{(\tau r)^{3}} [ f(\tau -r) (\tau^{2}
+r^{2})+ f'(\tau -r) \tau r(\tau -r)]
\end{eqnarray}
Clearly, these solutions are not related any more by a redefinition
of $f$ in one of them; this we see already from the different pole structure
for $f= \delta '$.

As a last example we start from a outgoing wave
$\phi_{3,1}=f(\tau -r)/r$ and proceed to $5+3$ dimensions:
\begin{eqnarray}
\phi_{5,3}(\tau ,r) &=&  - \frac{1}{\tau r^{3}} [r f''(\tau -r)
+ f'(\tau -r)] \\
\bar \phi_{5,3}(\tau ,r) &=& - \frac{1}{\tau r^{3}} [ f'(\tau -r) r (\tau -r)
- \tau f(\tau -r)]
\end{eqnarray}
If we choose $f = \delta $, then the above solution
$ \bar \phi_{5,3}$ vanishes.

The first prescription is more powerful, since the second one is restricted
to a simultaneous increase of spatial and temporal dimensions (hence it
does, e.g. not yield solutions in $1+q$ or $p+1$ dimensions). But the latter
is useful to complete the first one, because it provides basically different
solutions. We also note that the application of the two prescriptions
commutes. Hence if we proceed e.g. from a $\phi_{3,5}$ to $\bar \phi_{7,7}$,
it does not matter if we first go to $\phi_{5,5}$ and then to
$\bar \phi_{7,7}$, or if we start with $\bar \phi_{5,7}$ and then apply
the first prescription to obtain the same $\bar \phi_{7,7}$.
On the other hand, if we go e.g. from $\phi_{3,3}$ to a solution in
$7+7$ dimensions, we have to distinguish if we use the second rule
not at all, once or twice (regardless of the order of the operators).
If we start from the form (\ref{33}), then the solution in $7+7$
dimensions contains maximally 4, 3, 2 derivatives of $f$, respectively.

Generally: if we replace two applications of the first rule by one
application of the second rule, then the maximal pole strength
is reduced by one.

\section{Solutions, which depend only on $\xi \doteq (\tau -r)(\tau +r)$}

Here we treat the two signs between $\tau $ and $r$ democratically.
Let $\psi_{n}(\xi )$ be a solution in $n=p+q$ dimensions. In
\cite{jjg} it was noted that such solutions are:
\begin{equation} \label{xixi}
\psi_{2}(\xi ) = \ln \xi \ , \quad
\psi_{n}(\xi ) = \xi^{1-n/2} \qquad (n \geq 3)
\end{equation}
where for $n > 2$ we can not identify $p$ and $q$ separately.
Note that $\psi_{2}(\xi )$ is a special case of the form given
in (\ref{d13}).

This string of solutions corresponds to rule 1, since
\begin{equation} \label{partxi}
- \frac{1}{r} \partial_{r} \psi_{n}(\xi ) = 2 f'_{n}(\xi ) = \frac{1}{\tau}
\partial_{\tau} \psi_{n}(\xi )
\end{equation}
is indeed the solution given in (\ref{xixi}) for $n+2$.

The second rule is supposed to generate a different solution
for $n+4$, which would be strange since the set (\ref{xixi}) is complete
(up to constants).
But to look at (\ref{partxi}) we recognize that the second rule only
yields trivial solutions here: $\bar \psi_{n+4}(\xi ) \equiv 0$.
So for this special type of solutions only the first rule is useful.
For further details, see \cite{jjg}.

\section{Generalization to the Klein Gordon equation}

Now we generalize (\ref{we}) to the spherically symmetric Klein
Gordon equation, i.e. the relativistic equation of motion for
a {\em massive} scalar particle:
\begin{equation} \label{KG}
\Big[ \partial_{r}^{2} + \frac{p-1}{r} \partial_{r} - \partial_{\tau}^{2}
- \frac{q-1}{\tau} \partial_{\tau} + m^{2} \Big] \phi_{p,q}^{(m)}(\tau ,r) =0
\end{equation}
It is easy to check that both rules still hold for this generalized
case. The reason is simply that the factor $m^{2}$ commutes with the
differential operators. (However, additive constants are not arbitrary
any more in $\phi^{(m)}_{p,q}$.)

Again we search for solutions of the form $\psi^{(m)}_{n}(\xi )$.
In terms of $\xi $, eq. (\ref{KG}) reads:
\begin{equation} \label{KGxi}
\Big[ \xi \partial_{\xi}^{2} + \frac{n}{2} \partial_{\xi} - \Big(
\frac{m}{2} \Big) ^{2} \Big] \psi_{n}^{(m)}(\xi ) =0
\end{equation}
In \cite{GR}, p. 972 we find the solution of the differential eq.
\begin{equation}
u''(z) + \frac{1-\nu}{z} u'(z) - \frac{1}{4z} u(z) =0
\end{equation}
namely:
\begin{equation}
u(z) = z^{\nu /2} Z_{\nu}(i \sqrt{z})
\end{equation}
where $Z_{\nu}$ is any Bessel function.
For $m \neq 0$ we can transform (\ref{KGxi}) to this form by substituting
$z \doteq m^{2} \xi$.
Thus the solution is:
\begin{equation}
\psi^{(m)}_{n}(\xi ) = (m \sqrt{\xi})^{1-n/2} Z_{1-n/2}(im\sqrt{\xi })
\end{equation}
(where we assume $m>0$). In
\begin{equation}
\partial_{m^{2}} \psi^{(m)}_{n}(\xi ) = \frac{2-n}{4m^{2}} \psi^{(m)}_{n}
(\xi ) + \frac{1}{2} i \xi (m \sqrt{\xi})^{-n/2} Z'_{1-n/2}(im\sqrt{\xi})
\end{equation}
the derivated Bessel function can be expressed in various ways
in terms of (non derivated) $Z_{1-n/2^{+}_{-}1}$, which describes
the dynamics in different dimensions.

Of course, in the limit $m \to 0$ we recover the solutions of the
preceding section.

\section{Transition to even and fractional dimensions}

The first rule in the form of section 3 only permits steps
of two dimensions. To arrive at non odd dimensions from the explicit
solutions given above by means of that rule, we need a fractional
application of the operator, which provides the dimensional shift.
This concept becomes much simpler if we use for this operator
the identity:
\begin{equation}
-\frac{1}{r} \partial_{r} \equiv -2 \partial_{r^{2}}
\end{equation}
We ignore the factor -2 and postulate the natural {\em generalization of
rule 1}:

If  $\phi_{p,q}$ is a solution of the spherical wave eq. (\ref{we})
(or Klein Gordon eq. (\ref{KG})), then also
\begin{equation}
\phi_{p+2\alpha ,q} \doteq \partial_{r^{2}}^{\alpha} \phi_{p,q} \quad {\rm and}
\quad \phi_{p,q+2\beta} \doteq \partial_{\tau^{2}}^{\beta} \phi_{p,q}
\end{equation}
are solutions in $(p+2\alpha )+q$ and $p+(q+2\beta )$ dimensions,
respectively, for all $\alpha , \beta \in \reell $, if only $p+2\alpha
\geq 1 $ rsp. $q+2\beta \geq 1$.\\

We saw this before for integer $\alpha , \beta $, hence we just consider
non integers now.
Clearly the two statements are equivalent, so let's only consider
the first one.

For the fractional derivation -- or integration for $\alpha <0$ --
we refer to the definition
given by the Weyl transformation (see e.g. \cite{GS,Bateman})
\begin{eqnarray}
\phi_{p+2\alpha ,q}(\tau ,r) &=& \frac{1}{\Gamma (-\alpha )}
\int_{r^{2}}^{\infty} (u-r^{2})^{-1-\alpha} \phi_{p,q}(\tau ,\sqrt{u}) du \\
&=& \frac{1}{\Gamma (-\alpha )} \int_{0}^{\infty} x^{-1-\alpha}
\phi_{p,q}(\tau , \sqrt{r^{2}+x}) dx \nonumber
\end{eqnarray}
We substitute $y \doteq \sqrt{r^{2}+x}$, and we have to check now if
the following expression vanishes:
\begin{eqnarray*}
&& \Big[ \partial_{r}^{2} + \frac{p-1+2\alpha}{r} \partial_{r}
-\partial_{\tau}^{2} - \frac{q-1}{\tau} \partial_{\tau} \Big]
\int_{0}^{\infty} x^{-1-\alpha} \phi_{p,q}(\tau ,y) \\
& \propto & \int_{0}^{\infty} x^{-1-\alpha} \Big[ (1 - \frac{x}{y^{2}}
\partial_{y}^{2}+\frac{p+2\alpha - r^{2}/y^{2}}{y} \partial_{y}
- \partial_{\tau}^{2}- \frac{q-1}{\tau} \partial_{\tau} \Big] \phi_{p,q}
(\tau ,y) \\
&=& \int_{0}^{\infty}x^{-1-\alpha} \Big[ - \frac{2x}{y} \partial_{x}
+ \frac{2\alpha}{y} + \frac{x}{y^{3}} \Big] \partial_{y} \phi_{p,q}
(\tau ,y) dx
\end{eqnarray*}
In the sense of analytic continuation we can perform partial integration
in the first term without boundary terms. This shows immediately that this
expression vanishes, as we postulated. \\

In principle also the second rule could be generalized accordingly,
but the resulting fractional operator is very uneasy to handle.

Now we have a key for the construction of solutions in non odd dimensions.
General solution are for instance:
\begin{eqnarray}
\phi_{2,1}(t,r) &=& \int_{0}^{\infty}x^{-3/2}[F_{1}(t-\sqrt{r^{2}+x})+
F_{2}(t+\sqrt{r^{2}+x})] dx \\
\phi_{2,2}(\tau ,r) &=& \int _{0}^{\infty}dx \int_{0}^{\infty}dy (xy)^{-3/2}
\cdot \nonumber \\
&& [F_{1}(\sqrt{\tau^{2}+y}-\sqrt{r^{2}+x}) + F_{2}(\sqrt{\tau^{2}+y}+
\sqrt{r^{2}+x})]
\end{eqnarray}
etc.

Of course we would like to have really explicit solutions. But before
evaluating them for particular functions $F_{1},F_{2}$,
we construct some solutions for even dimensions in an independent way.

\section{Explicit solutions for even dimensions}

In particular the case of $2+1$ dimensions is motivated from solid
state physics: some crystals consist of layers, where the interaction
between layers is strongly suppressed with respect to the interactions
inside the layers.

In view of our recursion rules for the generation of solutions in higher
dimensions, we have to concentrate on finding particular solutions
for $2+1, \ 1+2 $ and $2+2$ dimensions.

So far, we only have the -- manifestly Lorentz invariant --
solutions of section 4 for even dimensions. Now we want to
construct different ones, and then discuss them in the context of
section 6.\\

If we look for polynomials in $r, \tau$ fulfilling (\ref{we}),
the only non trivial -- i.e. non constant -- solution is (except for
the linear solutions for $\phi_{1,1}$)
\begin{equation} \label{poly}
\phi_{p,p}(\tau ,r) = \tau^{2}+r^{2}
\end{equation}
So we have a new solution for any $p=q$, e.g. $\phi_{2,2}$, but it
does not yield a non trivial string of solutions from our theorems.
However, it is new, even in $p=q=1$, and it will help us to find
less obvious solutions.

We make an ansatz that generalizes somehow the solutions depending
only on $\xi $:
\begin{equation}
\phi (\tau ,r) = f(\tau ,r) (\tau^{2} \ ^{+}_{-} \ r^{2})^{\alpha}
\qquad (\alpha \in \reell )
\end{equation}
For the upper sign we can only reproduce (\ref{poly}), so we concentrate
on the lower sign (Lorentz invariant bracket), assume $\tau^{2} \neq r^{2}$
(off light cone) and denote $f_{r} \doteq \partial_{r}f$ etc.
We arrive at the condition:
\begin{equation} \label{fwe}
(\tau^{2}-r^{2}) \Big\{ f_{rr}+ \frac{p-1}{r}f_{r}-
f_{\tau \tau}-\frac{q-1}{\tau} f_{\tau} \Big\} = 4\alpha \Big\{
rf_{r}+\tau f_{\tau}+(\alpha -1 +\frac{n}{2}) f \Big\}
\end{equation}
(still with $n=p+q$). We want the curly brackets to vanish, i.e.
$f$ should also obey the wave eq. If we take the trivial solution
$f \equiv const.$, we obtain again the solutions (\ref{xixi}), so
we look for something more original.

Let's consider $q=1$ or $p=1$: then a simple solution of (\ref{fwe})
is $f=\tau $ rsp. $f=r$ and $\alpha =n/2$ and yields:
\begin{equation} \label{mix}
\phi_{p,1} = \frac{\tau}{(\tau^{2}-r^{2})^{(p+1)/2}} \quad ; \qquad
\phi_{1,q} = \frac{r}{(\tau^{2}-r^{2})^{(q+1)/2}}
\end{equation}
In $p=q=1$ we get special cases of the form (\ref{d13}), but more
interesting are $\phi_{2,1}$ and $\phi_{1,2}$ given in (\ref{mix}).
If we proceed by means of rule 1 to $\phi_{4,1}, \ \phi_{1,4}$
etc. we stay inside the set described by (\ref{mix}). But in addition
we obtain solutions for {\em all mixed} $p$ and $q$ (with respect
to even/odd), e.g.
\begin{equation}
\phi_{2,3} = \frac{1}{\tau} \partial_{\tau} \frac{\tau}{\tau^{2}-r^{2}}
= \frac{1}{\tau (\tau^{2}-r^{2})}- \frac{2\tau}{(\tau^{2}-r^{2})^{2}}
\end{equation}
etc.

We should still find new solutions for $p$ and $q$ both even.
The above solutions for $f$  fails (it leads to a trivial $\phi_{p,q}$),
therefore we try the polynomial solution: $f=\tau^{2}+r^{2}$ for $p=q$.
We obtain:
\begin{equation} \label{even}
\phi_{p,p} = \frac{\tau^{2}+r^{2}}{(\tau^{2}-r^{2})^{1+p}}
\end{equation}
Now we have in particular the desired solution for $\phi_{2,2}$ and
thus a string of explicit solutions for all even $p$ and $q$.
If we use the first rule to produce $\phi_{p+2,p+2}$ etc.,
we keep the form of (\ref{even}), and the second rule leads back to
the form (\ref{xixi}). But we can generate new solutions for $p \neq q$,
such as
\begin{equation}
\phi_{4,2} = \frac{2\tau^{2}+r^{2}}{(\tau^{2}-r^{2})^{4}} \quad ; \qquad
\phi_{2,4} = \frac{\tau^{2}+2r^{2}}{(\tau^{2}-r^{2})^{4}}
\end{equation}

Last we want to relate these solutions to the concept of
fractional derivatives described in section 6. If we start
from $\phi_{1,1}= \delta (\tau -r)$, then
\begin{equation}
\phi_{2,1}=\int_{0}^{\infty} x^{-3/2}\delta (\tau - \sqrt{r^{2}+x})dx
= \vert \tau \vert \int_{0}^{\infty} x^{-3/2}\delta (x-\tau^{2}+r^{2})
\end{equation}
If we drop the proportionality constant (including sign$(\tau )$),
we recover $\phi_{2,1}$ given in (\ref{mix}).

If we begin with $\phi_{1,1}=\Theta (\tau -r)$ or $\phi_{3,1}=
\delta (\tau -r)/r$ and move to $\phi_{2,1}$ by half a derivation
rsp. integration with respect to $r^{2}$, then we end up with the
form (\ref{xixi}). If we proceed from this solution to
$\phi_{2,2}$ by applying $\partial_{\tau^{2}}^{1/2}$, so we are still
in the set of solutions (\ref{xixi}).

\section{Construction of new solutions for fixed $p$ and $q$}

Up to now we considered procedures to build up solutions
in different dimensions from one given solution. Now that we
already know solutions for all integer $p+q$, we look for
different solutions in the same dimension as the given one.

For this we need an operator $A$, which commutes with the spherical
D'Alembert operator:
\begin{equation}
[(\partial_{r}^{2}+\frac{p-1}{r}\partial_{r}-\partial_{\tau}^{2}
-\frac{q-1}{\tau}\partial_{\tau}),A] = 0
\end{equation}
(Of course, $A=const.$ is not interesting.)

{\em\bf Rule 3} For $p=1$ rsp. $q=1$
the operator $A=\partial_{r}^{\alpha}$ rsp. $\partial_{\tau}^{\alpha}$
works for all $\alpha \in \reell^{+} $.

Consider as an example $\phi_{2,1} = (\tau^{2}-r^{2})^{-1/2}$ (included
in (\ref{xixi})). One derivation by $\tau $ yields immediately the solution
given in (\ref{mix}), and in addition we obtain:
\begin{equation}
\tilde \phi_{2,1}(\tau ,r) =
\partial_{\tau}^{2}(\tau^{2}-r^{2})^{-1/2} \propto \frac{2\tau^{2}+r^{2}}
{(\tau^{2}-r^{2})^{5/2}} \qquad {\rm etc.}
\end{equation}

But if the dimension, which is $>1$, is also odd, we don't win anything
because there we already have solutions involving arbitrary functions
of $(\tau \ ^{+}_{-} \ r)$. Take e.g. $\phi_{3,1}$ from (\ref{d13}):
application of rule 3 just reproduces the same structure.\\

What can we do if $p$ and $q$ are both $>1$ ? It can be seen easily
that any ansatz $A=P_{1}(\tau ,r) P_{2}(\partial_{\tau },
\partial_{r})$, where $P_{1},P_{2}$ are finite polynomials, fails.

But we can combine the generalized rule 1 with rule 3 to
obtain:

{\em\bf Rule 4} If $\phi_{p,q}$ is a solution, then also
\begin{displaymath}
\partial_{r^{2}}^{(p-1)/2} \partial_{r}^{\alpha} \partial_{r^{2}}
^{-(p-1)/2} \phi_{p,q} \qquad {\rm and} \qquad
\partial_{\tau^{2}}^{(q-1)/2} \partial_{\tau}^{\beta} \partial_{\tau^{2}}
^{-(q-1)/2} \phi_{p,q}
\end{displaymath}
are solutions in $p+q$ dimensions, for any $\alpha , \beta \in \reell^{+} $.\\

As an example we consider $\phi_{3,2}$ from (\ref{xixi}). For
$\alpha = 1,2 \dots $ we get further solutions in $3+2$ dimensions,
namely:
\begin{equation}
\frac{\tau^{2}/r+2r}{(\tau^{2}-r^{2})^{5/2}} \qquad , \qquad
\frac{3\tau^{2} + 2 r^{2}}{(\tau^{2}-r^{2})^{7/2}} \qquad {\rm etc.}
\end{equation}
Note, however, that the practical application of this theorem is often
complicated. In particular if $p$ and $q$ are both even, we are forced
to use fractional integration and derivation.\\

We may also build new solutions by a combination of the first and
second rule, which is different from unity, e.g. with the operator:
\begin{equation}
[\partial_{\tau^{2}} + \partial_{r^{2}}][\partial_{\tau^{2}} \partial_{r^{2}}
]^{-1} = \partial_{\tau^{2}}^{-1} + \partial_{r^{2}}^{-1}
\end{equation}

\section{Definition of $ \Box ^{\lambda}$ for $q>1$}

In \cite{BG,jjg} the analytic continuation of $(\tau^{2}-r^{2})^{\lambda}$
with respect to real $\lambda $ was discussed and applied extensively
in $d+1$ dimensions. The discussion revealed that its singularities
and residues are strongly related to physical properties. In particular
in \cite{jjg} it was shown that a displacement in dimensions, not
necessarily integers, can be interpreted as solutions for different
powers of the D'Alembert operator.

Now also this consideration shall be generalized to $p+q$ dimensions.
For that purpose, we need a definition of $\Box^{\lambda}$ for
multiple times. The extension of the analytic properties is not
straightforward, but displays interesting and qualitatively new
properties, as we will see.

Let us summarize the analytic properties of the distributions
\begin{displaymath}
P_{+-}^{\lambda} \doteq (\tau^{2} -r^{2})^{\lambda}_{+-} \quad {\rm and}
\quad (P+i0)^{\lambda}
\end{displaymath}
(see \cite{GS} p. 350 ff.). $+-$ means zero outside, inside the light cone,
respectively and $(P+i0)^{\lambda}$ means the limit $( \tau^{2} -r^{2}
+i\varepsilon )^{\lambda}$ when $\varepsilon \to 0$. The main results
are the following.\\

a) $p$ odd, $q$ even or v.v. \\
$P^{\lambda}_{+}$ has simple poles at $\lambda = -1, \ -2 , \dots $
and $\lambda = - n/2, \ -n/2-1, \dots $, where $n \doteq p+q$.
For $k \in \natur $ the residues are:
\begin{eqnarray}
^{res}_{\lambda \to -k} P^{\lambda}_{+} &=& \frac{(-1)^{k-1}}{(k-1)!}
\delta^{(k-1)}(P) \\
^{res}_{\lambda \to -n/2-k} P_{+}^{\lambda} &=& \frac{(-1)^{p/2}\pi^{n/2}}
{4^{k} k! \Gamma (n/2+k)} \Box^{k} \delta (x_{1}, \dots ,x_{n})
\end{eqnarray}
(here the components $x_{i}$ run over both types of coordinates).\\

b) $p,q$ even \\
$P^{\lambda}_{+}$ has simple poles for $\lambda = -1,-2, \dots ,-k,\dots $
\begin{eqnarray}
^{res}_{\lambda \to -k} P^{\lambda}_{+} &=& (-1)^{k-1} \delta^{(k-1)}(P)
\qquad \quad (k \leq n/2) \\
^{res}_{\lambda \to -n/2-k} P^{\lambda}_{+} &=& \frac{(-1)^{n/2+k-1}}
{(n/2+k-1)!} \delta^{(n/2+k-1)}(P) + \frac{(-1)^{p/2} \pi^{n/2}}{4^{k} k!
\Gamma (n/2+k)} \Box^{k} \delta (x_{1}, \dots ,x_{n})
\end{eqnarray}

c) $p,q$ odd \\
$P_{+}^{\lambda}$ has simple poles for $k=-1,-2 \dots ,-n/2+1$ with
\begin{equation}
^{res}_{\lambda \to -k} P_{+}^{\lambda} = \frac{(-1)^{k-1}}{(k-1)!}
\delta^{(k-1)}(P)
\end{equation}
and single {\em and double} poles for $\lambda = -n/2, -n/2-1, \dots $
with (observe the difference to $p,q$ even)
\begin{equation}
^{res}_{\lambda \to -n/2-k} P^{\lambda}_{+}= c_{1} \frac{ \Box^{k}
\delta (P)}{(\lambda+n/2+k)^{2}} + \frac{c_{2} \Box^{k} \delta
(x_{1}, \dots ,x_{n}) + c_{2}' \delta^{(n/2+k-1)}(P)}{ \lambda +n/2+k }
\end{equation}

For $P_{-}$ exchange $p$ and $q$  and replace $\delta^{(k-1)}(P)$ by
$\delta^{(k-1)}(-P)$ and $\Box $ by $-\Box $. \\

d) The poles of $(P+i0)^{\lambda}$  are simple for $\lambda = -n/2-k, \
k=0,1,2 \dots $ and
\begin{equation}
^{res}_{\lambda \to -n/2-k} (P+i0)^{\lambda} = \frac{e^{-i \pi p/2}
\pi^{n/2}}{4^{k} k! \Gamma (n/2+k)} \Box^{k} \delta(x_{1}, \dots ,x_{n})
\end{equation}
and its complex conjugate for $(P-i0)^{\lambda}$.\\

In \cite{BG} a definition of $\Box^{\lambda}$ for real $\lambda $ was given,
which reduces to $\Box^{k}$ if $\lambda \in \natur $. This was achieved
as follows
\begin{equation} \label{bogaeq}
\Box^{\lambda}_{R,A} * \phi (x)= \frac{2^{2\alpha +1}(t^{2}-r^{2})
^{-\alpha-n/2} \Theta (-+ t)}{\pi^{n/2-1} \Gamma(1-\alpha-n/2)
\Gamma(-\alpha )} * \phi (x)
\end{equation}
$R,A$ stand for retarded, advanced and refer to the negative, positive sign
in the argument of the step function $\Theta $.
It is easy to verify that (\ref{bogaeq})
reduces to $\Box^{k} \phi (x)$ when $\lambda =k$ and $q=1$, due to the
properties a), b) and c).

This is generally the case when $p,q$ are both odd. E.g. in dimensions
3+3 or 3+1 the double pole plays an essential role. (In 3+3 $t$ is to
be understood as $^{+}_{-} \tau $.)

So for a classical theory with $p,q$ both odd, $\Box^{\lambda}$ given
by (\ref{bogaeq}) is well defined, retarded as well as advanced.\\

The same happens if we consider a quantum theory, where (see \cite{BG})
\begin{equation} \label{bgquant}
\Box^{\lambda}_{^{+}_{-}} * \phi (x) = ^{+}_{-} i e^{^{+}_{-}i \pi (\lambda
+n/2)}4^{\lambda} (4\pi )^{n/2} \frac{\Gamma (\lambda +n/2)}
{\Gamma (- \lambda )}
(t^{2}-r^{2} {^{+}_{-}} i0)^{\lambda -n/2} * \phi (x)
\end{equation}
Thus if $\lambda = k \in \natur $ the numerator as well as the denominator
pick up a pole, and we are left with the residue $\Box^{k} \delta $
(see property d)). \\

If we try to repeat the same procedure for $p,q$ even, e.g. 4+2 dimensions,
the classical theory as described by eq. (\ref{bogaeq}), runs into
trouble. The above reasoning doesn't work any more, since $(t^{2}-r^{2})
_{+}^{\lambda}$ has only single poles, and there is no way to compensate
the double poles in the denominator occurring in (\ref{bogaeq}).
Everything would be swept away.

And even if we write a phenomenological $\Box^{\lambda}$ with a single
pole in the denominator, we arrive for integer $\lambda $ at a linear
combination
\begin{displaymath}
a_{1} \Box^{k} + a_{2} \delta^{(k-1)}(P)
\end{displaymath}
and {\em not} the desired result $\Box^{k}$.

However, this is not the case for a quantum theory of the form (\ref{bgquant}).
The latter leads in fact to $\Box^{k}$ for $\lambda \to k$.
Hence everything works for the quantum operator $\Box ^{\lambda}_{^{+}_{-}}$
with multiple times when $p$ and $q$ are both even or both odd, whereas
the classical definition for $\Box^{\lambda}$ fails for even dimensions.

So if we consider the quantum definition of $\Box^{\lambda}$, we can extend
the result of \cite{BG}, section V, according to which a derivation with
respect to $r^{2} \ (\tau^{2})$ increases the spatial (temporal) dimension
by 2 or diminishes the power $\alpha $ by 1. If we apply the operator
$\partial_{r^{2}}^{\gamma} \ (\partial_{\tau^{2}}^{\gamma})$ on a radial
solution, we increase $p$ ($q$) by $2\gamma $ or diminish $\alpha $ by
$\gamma $.

The results for the ``mixed case'' (with respect to even/odd) are obvious
and not plagued by any problems due to double poles.

\section{The wave equation in curved space}

In particular, the previous formalism can be used to transform a time
coordinate into a space coordinate. The operator, which does this job,
is
\begin{equation} \label{tr}
\partial_{\tau^{2}}^{-1/2} \ \partial_{r^{2}}^{1/2}
\end{equation}
Of course, this reminds us of the process taking place when we
enter a black hole. When crossing the Schwarzschild radius,
a time coordinate becomes spatial and vice versa, as we see
from the Schwarzschild metrics in polar coordinates:
\begin{equation} \label{Schwarzschild}
ds^{2}= \Big( 1-\frac{2M}{r} \Big) d\tau^{2} + \tau^{2} d \Omega_{q}^{2}
- \frac{1}{1-2M/r}dr^{2}-r^{2} d \Omega_{p}^{2}
\end{equation}
We refer to a spherically symmetric, static black hole solution.
$d\Omega_{q}, \ d\Omega_{p}$ are the surface elements on the temporal,
spatial unit sphere, respectively.
\footnote{The appearing singularity of the metrics {\em on} the boundary is an
artifact of the choice of the coordinates and can be cured by
a different choice, as Kruskal showed for 3+1 dimensions \cite{Kruskal}.}
(Generally, a sensible dimensional
continuation of general relativity is given by the Lovelock eqs,
\cite{Lovelock}. For its application on static black holes,
see \cite{DBH}.)

Formula (\ref{Schwarzschild}) displays immediately that in a transition
from the outside to the inside of a black hole (of radius $2M$)
the radial temporal component $\tau $ becomes spatial while the radial
spatial component $r$ becomes temporal. The character of the
further components remains unchanged, so we have a transition
$p+q \to p+q$. (However, inside the black hole, the angular terms in
(\ref{Schwarzschild}) suffer from a mismatch between the radial and
the angular factor.)

Let's consider this in view of spherically symmetric waves.
The time-to-space transition we might describe with the operator (\ref{tr}),
but to include also
the simultaneous space-to-time transition we have no reasonable alternative
to its inverse operator, so we don't end up with anything instructive.

We could describe a transition of $3+1 \to 3+1$ -- or generally:
$p+q \to p+q$ -- dimensions different from unity, e.g. by using
the procedures of section 8,
but a motivation for this remains to be found.\\

A highly interesting transition, which is {\em not} $p+q \to p+q$ is the
{\em Wick rotation}. It transforms e.g. Schr\"{o}dinger's eq. to the
diffusion eq., hence deterministic and reversible quantum mechanics
to an irreversible stochastic process (Brownian motion).

But unfortunately our transformation rules 1 and 2 fail as soon as
$p$ or $q$ vanishes, since we have always assumed $\partial_{r}^{2}-
\partial_{\tau}^{2}$ to be part of the D'Alembert operator. In
Euclidean space we would be left with the search for harmonic functions,
which is well established in the literature.\\

Still, the most attractive feature in this context seems to us the
exchange of a time against a space coordinate in gravitation theory.
But in order to approach such questions as the transition
across the boundary of black holes seriously,
we  have to expand our discussion to a curved space.

We add some simple remarks on this generalization. However, much
of this extension remains to be worked out. Here we want
to illustrate one interesting property, which is related to
the previous discussion of dimensional shifts. As we will see,
certain types of curvatures can be described in a flat metrics
by altering the dimensions.

Assume the
temporal and spatial sector of the metrics to be decoupled:
\begin{equation} \label{ggg}
g= \left( \begin{array}{cc} g^{(\tau )} & 0 \\ 0 & g^{(r)} \end{array}
\right) \ ,
\end{equation}
where $g^{(\tau )}, \ g^{(r)}$ is a $q \times q , \ p\times p$ matrix,
respectively.

First we consider the spatial part of the generalized D'Alembert
operator in this metrics, which is the Laplace-Beltrami operator:
\begin{equation}  \label{LB}
\Delta = \frac{1}{\sqrt{\vert det \ g^{(r)}\vert }} \partial_{\mu} \Big[
\sqrt{\vert det \ g^{(r)}\vert } \ g^{(r) \mu \nu} \Big] \partial_{\nu}
\end{equation}
In flat space and polar coordinates $r, \theta_{1} \dots \theta_{p-1}$
we have $g^{(r)rr}=1$ and
$\sqrt{\vert det \ g^{(r)} \vert } = r^{p-1} \sin^{p-2}\theta_{p-1}
\sin^{p-3}\theta_{p-2} \dots \sin \theta_{2}$.
Even if we generalize this to
\begin{equation}
\sqrt{\vert det g^{(r)} \vert }= r^{p-1} F(\theta_{1}, \dots ,\theta_{p-1}) \ ,
\end{equation}
where $F$ is an arbitrary function, we always end up with the radial part
\begin{equation}
\Delta_{(r)} = \partial_{r}^{2} + \frac{p-1}{r} \partial_{r}
\end{equation}
that we have used so far.

A generalization of $\Delta_{(r)}$ can be achieved, however, if we let
the matrix elements $g^{(r)\mu \nu}$ depend on $r$. We still consider
the spatial part separately and assume that it takes the form:
\begin{equation}
(g^{(r)\mu \nu}) = \left( \begin{array}{cccc} g^{rr} & 0 & \dots & 0 \\
0 &&& \\ : & & g_{ij} & \\ 0 &&& \\ \end{array} \right)
\end{equation}
with $i,j=1, \dots ,p-1$, i.e. in addition the radial and the
angular part are decoupled. Hence:
\begin{equation} \label{detg}
det g^{(r)} = g^{rr} \cdot det (g_{ij})
\end{equation}
Let us consider the general case where both factors in (\ref{detg})
pick up a non trivial dependence on $r$.
\begin{eqnarray} \label{met1}
g^{rr} &=& f_{1}(r) \\
det (g_{ij}) &=& f_{2}(r) r^{p-1} F(\phi_{i}) \label{met2}
\end{eqnarray}
where $f_{1},f_{2}$ are functions $\reell ^{+} \to \reell ^{+}$.
Inserting this into (\ref{LB}) we obtain:
\begin{equation} \label{delr}
\Delta_{(r)} = f_{1}(r) \left[ \partial^{2}_{r} + \Big\{
\frac{p-1}{r} + \frac{3}{2} ( \partial_{r} \ln f_{1}(r)) + \frac{1}{2}
(\partial_{r} \ln f_{2}(r)) \Big\} \partial_{r} \right]
\end{equation}
In general, such functions $f_{1},f_{2}$ require new types of solutions,
completely different from the case $f_{1}=f_{2}= const.$ considered
above. An exception is the case, where they take the simple monomial
form:
\begin{eqnarray}
f_{1}(r) &=& c_{1} r^{\alpha} \label{f1} \\
f_{2}(r) &=& c_{2} r^{\beta} \qquad
(c_{1},c_{2},\alpha , \beta : \ constants \in \reell ) \label{f2}
\end{eqnarray}
Here the modification of the dynamics due to the metrics -- with
respect to the flat space -- corresponds to a dimensional shift
in the flat space (as long as $c_{1},c_{2} \neq 0, \ 3\alpha
+\beta \geq 2(1-p)$):
\begin{equation}
p \to p + \frac{3}{2} \alpha + \frac{1}{2} \beta
\end{equation}
Hence ironing the curved space -- as it is very standard -- corresponds
for the special metrics given in (\ref{met1},\ref{met2},\ref{f1},\ref{f2})
-- to a shift in the flat space dimension.

Thus in a Euclidean space we have to look for harmonic functions in a modified
dimension. E.g. for critical phenomena it would be
most crucial if we could argue that the effective dimension of
the flat space, where most models are defined, is slightly
different from four, due to a weak curvature of this type.\\

Now we remember the temporal part, which we have ignored for
the moment. We still assume that the metrics does not couple it to
the spatial part, see (\ref{ggg}). We are interested now
in cases where the metrics causes a shift in the flat
dimensions $p$ and/or $q$, the same effect that was obtained
before from other operations.

If we want to leave the temporal part in the flat form, then
the factor $f_{1}(r)$ in (\ref{delr}) disturbs in the sense
that we do not arrive at the flat D'Alembert operator in a
modified dimension unless $f_{1}(r) \equiv 1$. For different
functions $f_{1}$, the flat solutions are only valid if they are
separable: $\phi_{p,q}(\tau ,r) = \psi_{1}(\tau) \psi_{2}(r)$;
$\psi_{1},\psi_{2}$ being harmonic functions in $p,q$ dimensions,
respectively. Also an angle dependent factor in (\ref{met1})
would disturb in this sense.

In order to use flat solutions which do not have
this separable form, we have to deal with
$f_{1}(r) \equiv 1$, e.g. by using Riemann normal coordinates,
and arrange the dimensional shift only
by a non constant $f_{2}(r)$ in (\ref{met2}) of the form (\ref{f2}).

Of course the analogous statements hold if we want to introduce
a curved temporal metrics and keep a flat spatial metrics.
If we only use curvatures of the type (\ref{met2},\ref{f2}), then we can
easily shift $p$ and $q$ simultaneously.\\

Finally, we can also cause a simultaneous
modification of the flat $p$ and $q$ by the following choice:
\begin{equation}
g^{(r)rr} = g^{(\tau)\tau \tau } = c_{1} r^{\alpha} \tau ^{\bar \alpha}
\qquad (c_{1},\alpha, \bar \alpha : \ constants \in \reell )
\end{equation}
If we combine this with
\begin{equation}
f_{2}(r) = c_{2} r^{\beta} \ , \quad \bar f_{2}(\tau) =
\bar c_{2} \tau^{\bar \beta}
\end{equation}
where $\bar f_{2}(\tau)$ is the temporal analogue to $f_{2}(r)$,
then we arrive at the dimensional transformation:
\begin{equation}
p \to  p + \frac{3}{2} \alpha + \frac{1}{2} \beta \quad , \qquad
q \to q + \frac{3}{2} \bar \alpha + \frac{1}{2} \bar \beta \quad .
\end{equation}

\section{Conclusions}

We have provided a reservoir of solutions of the spherically symmetric
wave equation in $p+q$ dimensions and gave some insight into their structure.
We gave a large number of explicit solutions and a set of prescriptions
for constructing new solutions out of them. In particular we showed how
to modify the dynamics such that it fulfills the wave eq. in different
temporal or spatial dimensions. The same prescriptions also hold for the
Klein Gordon equation.

The transition in steps of two dimensions is very simple. Its analytic
continuation allows also for transitions by one or by fractional
dimensions, but its application is somewhat more involved. Such transitions
in the flat space also correspond to certain types of curvature, i.e.
special curved metrics can be described in the flat space by altering
its dimensions.

We found an operator transforming a space into a time coordinate or
vice versa, a transition that actually takes place when we cross
the boundary of a black hole. However, our description of this process
if not complete yet.

The analytic continuation of $\Box^{\lambda}$ with respect to $\lambda $
corresponds in some cases again to the dynamics of the standard
D'Alembert operator in modified dimensions. This continuation is
feasible for the classical as well as for the quantum definition in
odd $p$ and/or odd $q$. If $p,q$ are {\em both} even, however,
the classical definition fails
and only the quantum definition yields a sensible result.\\

{\bf Acknowledgement} \ One of us (J.J.G.) is indebted to
O. Obregon for motivating discussions.

\end{document}